# Shear induced collective diffusivity in an emulsion of viscous drops using dynamic structure factor: effects of viscosity ratio

Abhilash Reddy Malipeddi and Kausik Sarkar

*Department of Mechanical and Aerospace Engineering*

*The George Washington University*

*Washington, D.C., U.S.A. – 20052*

## Abstract

The shear induced collective diffusivity in an emulsion of viscous drops, specifically as a function of viscosity ratio, was computed using a fully resolved numerical method. An initially randomly packed layer of viscous drops spreading due to drop-drop interactions in an imposed shear has been simulated. The shear induced collective diffusivity coefficient was computed using a self-similar solution of the drop concentration profile. We also directly obtained the collective diffusivity (the collective diffusivity coefficient multiplied by the average drop volume fraction) itself, computing the dynamic structure factor from the simulated drop positions—an analysis typically applied only to homogeneous systems. The two quantities computed using entirely different methods are in broad agreement including their predictions of nonmonotonic variations with increasing capillary number and viscosity ratio. The computed values were also found to roughly match with past experimental measurements. The gradient diffusivity coefficient computed here, as expected, was found to be roughly one order of magnitude larger than the self-diffusivity for a dilute emulsion previously computed using pair-wise simulation of viscous drops in shear. The gradient diffusivity was found to be non-monotonic with increasing capillary number, similar to what was found for the self-diffusivity computed previously. However, gradient diffusivity also showed non-monotonicity with increasing viscosity ratio, unlike the previously computed self-diffusivity. The difference in variation could arise from drops not reaching equilibrium deformation between interactions—an effect absent in the pair-wise simulation used for computation of self-diffusivity—or from an intrinsic difference in physics underlying the two diffusivities. Indeed we offer a qualitative explanation of the nonmonotonic variation by relating it to average nonmonotonic drop deformation with increasing viscosity ratio. We also provided empirical correlations of the collective diffusivity as a function of viscosity ratio and capillary number.

# 1 Introduction

Non-colloidal particles in a sheared suspension or emulsion undergo a diffusive motion due to shear induced hydrodynamic interactions between particles [1-5] which can be of great importance in chemical and biomedical flows. Specifically, for red blood cells (RBC) in blood vessels at the physical hematocrit level of ~45%, such shear induced diffusion plays a critical role in determining their interactions and spatial concentration [6, 7]. Recently, we computed collective or gradient diffusivity in a concentrated viscous emulsion for the first time using fully resolved direct numerical simulation for a viscosity matched system [8]. In this short article, we extend the analysis to emulsions where the drop viscosity differs from the matrix viscosity. Additional novelty of the work stems from an alternative means of computing collective diffusivity using dynamic structure factor theory in a non-homogeneous system. The dynamic structure factor theory has previously been used only in homogeneous systems.

For a review of the literature of the shear induced diffusion we refer to our recent paper [8]. Briefly, shear induced diffusion is characterized by self-diffusivity $D_s = \dot{\gamma} a^2 f_s(\phi)$ ($\dot{\gamma}$ is the shear rate, $a$ the particle radius, and $f_s(\phi)$ the non-dimensional self-diffusivity, $\phi$ volume fraction) that defines the random motion of individual particles and present even in a homogeneous suspension or emulsion, and collective or gradient diffusivity $D_c = \dot{\gamma} a^2 f_c(\phi)$ ($f_c(\phi)$ is the non-dimensional collective diffusivity) that defines the diffusive flux $-D_c \nabla \phi$ in presence of a concentration gradient [9]. Shear induced self- and collective diffusion in rigid sphere suspensions have been widely studied both experimentally [3, 10] and numerically [11-15] since the pioneering work of Eckstein, Bailey [1]. However, emulsions of viscous drops in contrast hasn't been studied much. The first measurement of collective diffusivity in a viscous emulsion by King and Leighton [16] was marred by presence of stabilizing surfactants and gave rise to values much smaller than theoretically expected. The only successful measurement of collective diffusivity in a viscous emulsion in the literature was performed by Hudson [17] who used more viscous drops to avoid emulsion instability. Self and collective diffusivities in RBCs and vesicles have been measured in *in vitro* channels [6, 7, 18, 19]. Self-diffusivity in a viscous emulsion has been measured using pair-interactions between drops [4] in a dilute system as well as using full scale simulation in a non-dilute system [20]. Pair-wise simulation of vesicles [21, 22] and RBCs [23] have been used to compute self-diffusivity in a dilute system of such complex rheological particles. Note that unlike self-diffusivity, collective diffusivity computation by summing pair-wise displacement results in divergent integrals. Such problems could be addressed by a renormalization procedure using global constraints, as has been applied in analytical computation of effective stresses and sedimentation velocity in a rigid sphere suspension [24, 25].

In our previous article, we proposed a technique to compute collective diffusivity by numerically simulating a layer of initially closely packed drops diffusing in a plane shear. We used front tracking finite difference method [26-29] to fully resolve each drop deforming and moving past each other. We obtained values for non-dimensional collective diffusivity coefficient that compared well with previous experiments [6, 17]. Our simulation shows a non-monotonic variation in collective diffusivity with capillary number where with increasing capillary number, initially the diffusivity increases, reaches a maximum and subsequently decreases for larger values of capillary number. Although there was no other study of collective diffusivity versus capillary number in the literature, the self-diffusivity values computed using pair-wise simulation by Loewenberg and Hinch [4] also showed a similar non-monotonic trend.

Our previous study of a viscosity-matched system is extended here to systems where drop and matrix viscosities differ. Note that due to its ubiquitous presence in many chemical and biological phenomena, and at the same time the difficulties in performing controlled experiments and the concomitant sparse literature till date, the shear induced diffusion warrants further fundamental studies, such as the one attempted here, to delineate the nature of shear induced diffusion. We have also developed a dynamic structure factor theory based method to compute collective diffusivity. This method has so far been limited to homogeneous system. Furthermore, the main result found here—the non-monotonic variation of collective diffusivity with viscosity ratio, which differs from variation of self-diffusivity obtained previously using pair-wise interactions [4]—a posteriori justifies the current study. We provide a detailed comparison with their pair-wise drop trajectories to establish the accuracy of our simulation technique.

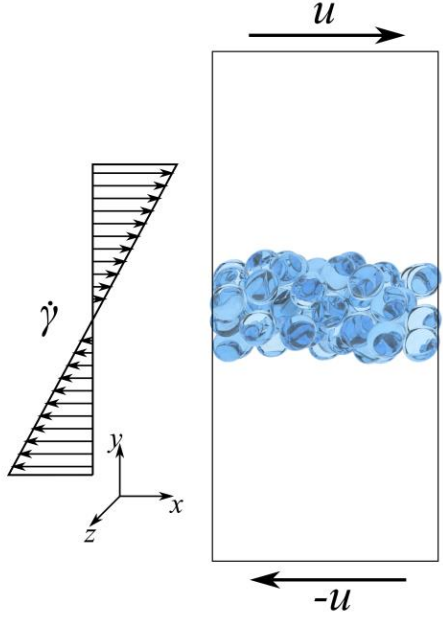

*Figure 1 A schematic of the layer of randomly placed drops in simple shear flow.*

## 2 Theoretical formulation

In this section, we provide the mathematical formulation and the numerical technique. The description here closely follows that of our previous article [8] and presented here briefly for completeness. A layer of randomly packed drops of radius $a$ and viscosity $\lambda_d$ suspended in a fluid of viscosity $\lambda_m$ is subjected to simple shear flow (Figure 1). They interact and move past each other resulting in effectively a diffusive motion.

### 2.1 Gradient diffusivity from self-similar solution

The problem (it is homogeneous in the $x$ and $z$ directions) results in a diffusion equation for the local volume fraction, $\phi = \phi(y,t)$ in the $y$ direction:

$$\frac{\partial \phi}{\partial t} = \frac{\partial}{\partial y}(D_c \frac{\partial \phi}{\partial y}), \qquad (1)$$

with $D_c = \dot{\gamma}\phi a^2 f_2$ (assumption of two-particle interactions being dominant and thereby the rate of collision being $\dot{\gamma}\phi$) [4-6, 30]. Here $f_2$ (subscript $c$ for collective diffusivity $f_{c,2}$ has been dropped for convenience) is the nondimensional collective diffusivity in the velocity-gradient direction, the focus of the present work.

The assumption of linear dependence of $D_c$ on volume fraction, or equivalently the dominance of pairwise interaction is *a posteriori* justified by the simulation results.

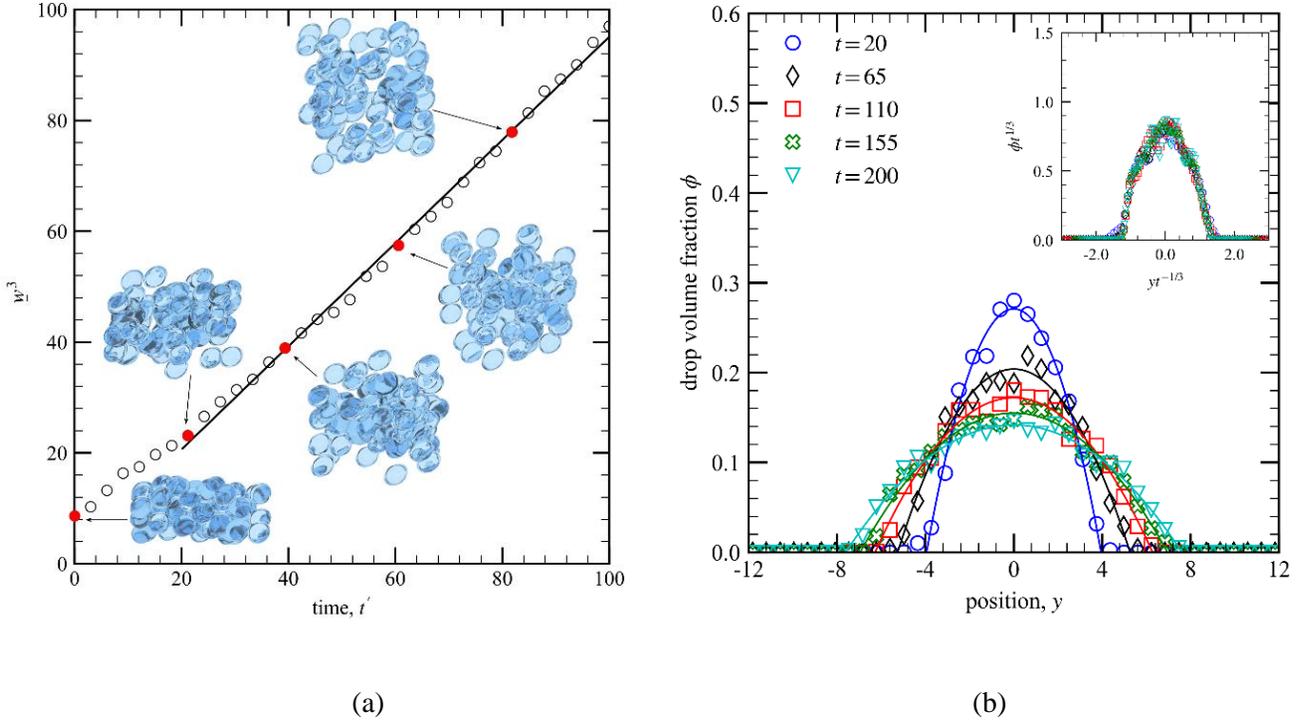

(a)                                    (b)

*Figure 2 (a) Cube of the width of the layer of drops is plotted as a function of time grows linearly with time. Snapshots of the evolving drop configurations at various time instants. (b) Concentration profile of the drops at various time instants, showing a parabolic profile broadening with time. Inset shows a collapse of the same when plotted against the scaled similarity variable according to Eq. (2).*

It has been shown by a detailed analysis that when a fixed number of particles spread due to shear induced diffusion, Equation (1), nondimensionalized using $t = t'/\dot{\gamma}, y = y'/a$. admits a self-similar parabolic concentration [6]

$$\psi(\eta) = (f_2 t')^{1/3} \phi = (b - \eta^2 / 6), \qquad \eta = y'/(f_2 t')^{1/3} \tag{2}$$

in the similarity variable $\eta$ ($b$ is a free parameter). Note the $t^{1/3}$ spread of the profile in contrast to $t^{1/2}$ growth in systems with a constant diffusivity $D_c$ (i.e. independent of volume fraction). Note that in case of particles spreading from one side in an initially Heaviside concentration profile [30], the characteristic exponent is still 1/2, even when $D_c = \dot{\gamma} \phi^m a^2 f_2$ with $m > 1$. It can be shown [6], that according to (2), the half-width $w$ of the $\phi(y')$ profile at half-height, satisfies

$$\underline{w}^3 - \underline{w}_o^3 = Kt', \qquad K = 9f_2 N_0 / (4\sqrt{2}), \qquad N_o = \int \phi(y',t')\,dy'. \qquad (3)$$

Here $\underline{w}_0$ is the initial width and $N_o$ is a conserved quantity, related to the particular nature of the problem mentioned before—a fixed number of particles diffusing out. $f_2$ is computed by fitting a parabolic curve for the droplet concentration at any instant of time and obtaining the half-width $\underline{w}$ (t'). In Malipeddi and Sarkar [8], we also proposed and developed an alternative method that avoids curve fitting, and computes the standard deviation, called "modified" width, of the concentration profile from the drop positions ($y_i'$) as

$$w = \sqrt{\frac{1}{N}\sum_{i=1}^{N}(y_i' - \mu)^2}, \quad \mu = \frac{1}{N}\sum_{i=1}^{N} y_i' \qquad (4)$$

Similar to Eq. 3, the cube of the modified width, $w^3$, scales linearly with time, albeit with a slightly different constant

$$w^3 - w_o^3 = K't', \qquad K' = 9f_2 N_0 / (10\sqrt{5}) \qquad (5)$$

In Malipeddi and Sarkar [8] we showed that the two procedures gave rise to identical values within statistical variations intrinsic to the system. Here, we use Eq. (5) to obtain the value of the collective diffusivity $f_2$. Each simulation is run till $t' = 200$. We discard the data in the initial transient region before the self-similar profile is reached (t<20). The remaining portion of the data is split into 4 smaller intervals of 45 inverse shear units. The length of these intervals is enough to ensure that the drop movements in each interval are uncorrelated. The slope of $w^3 - w_o^3$ vs time for each sub-interval is calculated and the mean value of these is reported. The standard deviation of the value across these sub-intervals is used to estimate the uncertainty in the measurement.

### 2.2 Gradient diffusivity from dynamic structure factor

Both self- and collective diffusivities can also be computed from particle dynamics in a homogeneous system [9, 13]. The computation of collective or gradient diffusivity—which measures diffusion down a concentration gradient—in a homogeneous system is counter-intuitive; it is predicted on the analysis of the decay of the spontaneously arising stochastic fluctuations encapsulated in the wave number dependent dynamic structure factor. Originally the theory stemmed from computation of diffusivity from dynamic light scattering (DLS), where scattered response of a monochromatic beam of laser from a scattering volume containing multiple scatterers (large macromolecules such as DNA, proteins, amino acids, viruses and bacteria) is measured [31]. For a dilute system of non-interacting scatterers, the autocorrelation of the fluctuation decays exponentially and the decay time is inversely proportional to the diffusivity. For concentrated systems, hydrodynamic interactions between the fluctuating particles cannot be neglected and the measured scattered response requires careful analysis [32-34] and proper interpretation. In different limits it reduces to collective- or self-diffusivity [9]. Leshansky and Brady [14] carefully described the theory and applied to shear induced diffusion. The analysis assumes no coalescence or breakup, as was also true for our simulation.

The scattered response at wavenumber $\mathbf{k}$ (nondimensionalized by $a$) from $N$ scatterers located at $\mathbf{x}'_\alpha(t'), \alpha = 1, 2, ... N$ is proportional to the intermediate scattering function

$$F(\mathbf{k}, t') = \frac{1}{N} \left\langle \sum_{\alpha, \beta = 1}^{N} e^{i\mathbf{k} \cdot (\mathbf{x}'_\alpha(t') - \mathbf{x}'_\beta(0))} \right\rangle. \tag{6}$$

Note that using the property of Dirac delta function, the number density of the scatterers (here droplets) and its spatial Fourier transform can be written as

$$n(\mathbf{x}', t') = \sum_{\alpha=1}^{N} \delta(\mathbf{x}' - \mathbf{x}'_\alpha) \qquad \hat{n}(\mathbf{k}, t') = \sum_{\alpha=1}^{N} e^{i\mathbf{k} \cdot \mathbf{x}'_\alpha}. \tag{7}$$

Therefore, $F(\mathbf{k}, t') = 1/N \langle \hat{n}(\mathbf{k}, t') \hat{n}^*(\mathbf{k}, 0) \rangle$ may be regarded as measuring the autocorrelation of the fluctuation $n'(\mathbf{x}', t')$ (where $n(\mathbf{x}', t) = n_0 + n'(\mathbf{x}', t')$) at wavenumber $\mathbf{k}$ for a statistically homogeneous system, as the constant background $n_0$ would not contribute to the autocorrelation. The system is not homogeneous but evolves from a nonhomogeneous initial condition. Leshansky and Brady [14] showed that the number density satisfies an advection diffusion equation in a shear flow $\mathbf{U} + \dot{\mathbf{\Gamma}} \cdot \mathbf{x}$ ($\mathbf{U}$ is the average flow and $\dot{\mathbf{\Gamma}}$ is the velocity gradient tensor):

$$\frac{\partial n}{\partial t} + (\mathbf{U} + \dot{\mathbf{\Gamma}} \cdot \mathbf{x}) \cdot \nabla n = \mathbf{D}^c \nabla^2 n. \tag{8}$$

In spite of the advection terms in equation (8), in a simple shear due to the orthogonality of the $\mathbf{k}$ ($= k\hat{\mathbf{y}}$) vector to the velocity field, one obtains a simple relation for the diffusivity in the gradient direction [14]:

$$D_{yy}^c = -\frac{1}{k^2}\frac{d(\ln F)}{dt'} \tag{9}$$

### 2.3 Direct numerical simulation

We solve the incompressible Navier-Stokes equations using a front-tracking method [27, 35]:

$$\nabla \cdot \mathbf{u} = 0,$$

$$\frac{\partial(\rho\mathbf{u})}{\partial t} + \nabla \cdot (\rho\mathbf{u}\mathbf{u}) = -\nabla p + \nabla \cdot \left[\mu\left\{\nabla\mathbf{u} + (\nabla\mathbf{u})^T\right\}\right] - \int_{\partial B} \kappa\mathbf{n}\sigma\delta(\mathbf{x}-\mathbf{x}')dS(\mathbf{x}'). \tag{10}$$

Here **u**, $p$, $\rho$ and $\mu$ are the velocity, pressure, density and viscosity respectively. $\kappa$ is the local drop surface curvature, **n** is the unit outward normal to the surface $\partial B$ of all drops and $\sigma$ is the interfacial tension. In this numerical method, all drops, and their interactions are resolved. The method has been used by our group in many problems involving drops [27, 36, 37] and capsules [38-40] in viscous and viscoelastic fluids [41-48]. Here, a uniform shear flow is generated in a computational domain, which is periodic in the $x$ and $z$ directions and has numerical walls in the $y$ direction moving with specified velocities (Figure 1). The distance between the walls is $L_y = 28a$ ($a$ is the drop radius) sufficiently large to simulate an unbounded shear. The length of the domain in the $x$ and $z$ directions is $L_x = L_z = 14a$. A 96×192×96 uniform grid is used in the computational domain leading to 15 grid points per drop diameter, shown to be sufficient in our earlier studies. In our previous article [8], we carefully varied the domain lengths to ensure that the results are independent of them. There we also showed that the results didn't change with drop numbers above 70 or correspondingly with $N_0$ (=1.43 here) beyond a value. The simulation here are executed with 70 drops.

## 3 Results and Discussion

### 3.1 Effects of viscosity ratio variation

We calculate the collective diffusivity coefficient $f_2$ for a range of viscosity ratio using the self-similar scaling of the drop layer width. We also compute the collective diffusivity itself $D_{yy}^c$ using the wave-number dependent dynamic structure factor (DSF).

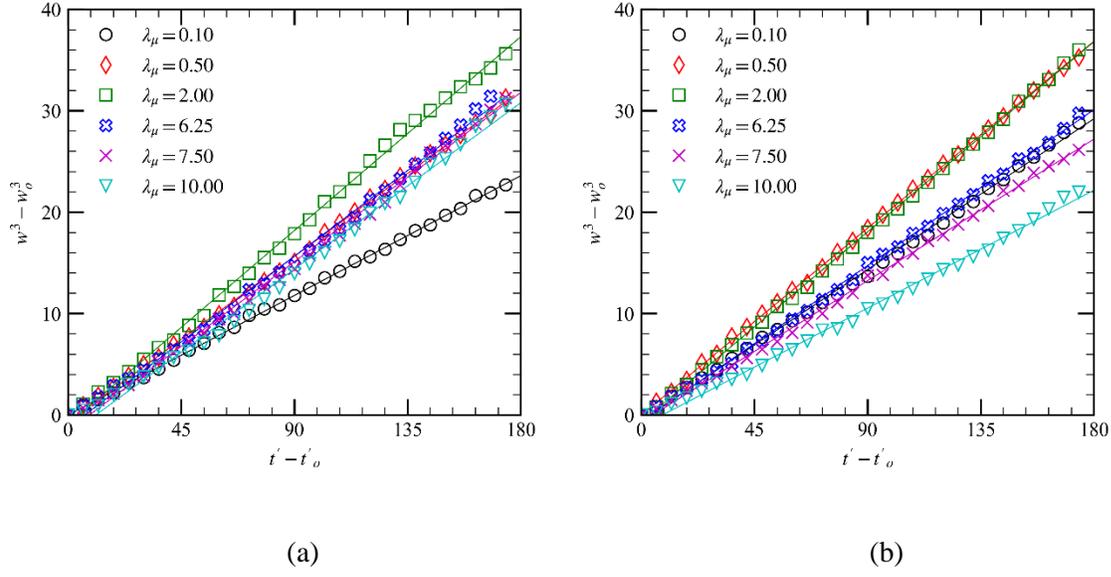

*Figure 3 Cube of Width of the drop layer vs time, for various viscosity ratios for Ca=0.05(a) and Ca=0.30(b) showing the expected linear trend.*

### 3.1.1  $f_2$ from scaling of the layer width

Figure 3 shows the growth of the width of the drop layer with time for different viscosity ratios at $Ca =0.05$ (Figure 3a) and $Ca =0.30$ (Figure 3b) displaying clearly a $1/3^{rd}$ scaling of the layer width with time. From the slopes, $f_2$ is computed according to Eq.(3), and plotted in Figure 5(a). As can be expected form Figure 3, slope varies non-monotonically with $\lambda_\mu$ (Figure 4a): $f_2$ increases initially with increasing $\lambda_\mu$ and reaches a maximum around the viscosity ratio of about 1, and then decreases. At higher viscosity ratios, drop deformation is known to decrease, reaching the limit of a rigid sphere, thereby resulting in lower diffusivity. In fact due to reversibility of the Stokes flow, pair-wise interaction fails to produce any diffusive motion in the rigid-particle limit. In a rigid-sphere suspension, $D_{yy} \sim \phi^2$ in the leading order [5], and therefore $f_2 = 0$.

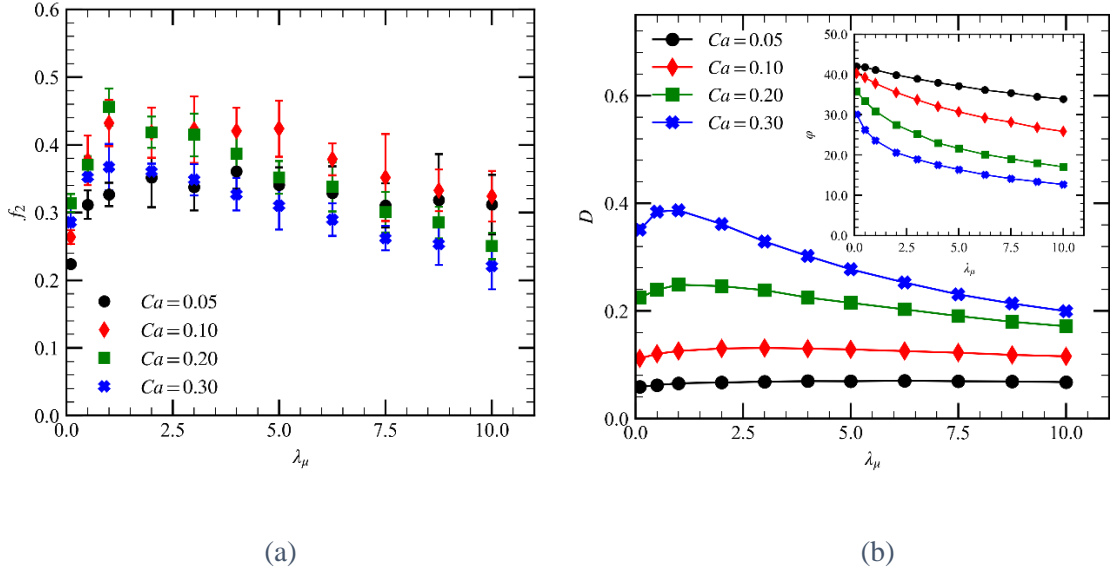

*Figure 4 (a) Dimensionless coefficient of diffusion, $f_2$ vs. viscosity ratio, $\lambda_\mu$, for different capillary numbers. (b) Main figure shows mean drop deformation parameter and inset shows drop orientation angle.*

Note that we found a nonmonotonic variation $f_2$ with $Ca$ previously [8]. Similar to the case there, the reason for the nonmonotonic behavior is the drop geometry as a function of $\lambda_\mu$. In Figure 4(b), we compute drop deformation characterized by Taylor deformation $D = (L-B)/(L+B)$ [$L$ being the maximum distance of the drop interface from its center and the $B$ being the minimum distance] [49], averaged over all drops, once it reaches a steady value after initial transients. It shows that with increasing $\lambda_\mu$, $D$ first increases, increase roughly coinciding with the region where $f_2$ also increases, and then decreases. At the same time the drop inclination angle $\varphi$, also plotted in Figure 4(b) with the flow direction steadily decreases. The increasing trend of $D$ and the variation of $\varphi$ are consistent with the small deformation moderate $\lambda_\mu$ perturbative results $D = Ca(19\lambda+16)/(16\lambda+16)$ [49] and $\theta = \pi/4 - Ca(2\lambda+3)(19\lambda+16)/(80\lambda+80)$ [50] [As is well recognized in the literature, the analytical result fails to predict decreasing trend of $D$ at higher viscosity ratio]. The initially increasing drop deformation increases hindrance to passage of drops past each other, increasing diffusivity coefficient $f_2$. Subsequently the decreasing deformation as well as decreasing inclination angle facilitates drops passing each other, reducing $f_2$. Note that Rusconi and Stone [30] have shown that the geometry of the suspended particles can have significant effects on gradient diffusivity, e.g., highly asymmetric plate-like particles have gradient diffusivity two orders of magnitude higher than that of rigid spheres at similar volume fraction.

We find collective diffusivity coefficient computed here as well as in our previous paper [8] consistently 8-9 times higher than the ones computed using simulation of pair-interaction between drops for a dilute emulsion by Loewenberg and Hinch [4]. This ratio is similar to what was found for rough sphere suspensions—ratio of collective diffusivity to self-diffusivity is ~6 [5]. As noted before Loewenberg and Hinch [4] found self-diffusivity to vary non-monotonically with increasing $Ca$ as did we [8], (also see below). However, their self-diffusivity coefficient $f_{s,2}$ didn't show a non-monotonic variation with $\lambda_\mu$; it showed a constant variation for the small $\lambda_\mu$ and then a strictly monotonic decay for larger values of $\lambda_\mu$. Using an identical approach of simulating pair-interactions, Omori, Ishikawa [23] computed the self-diffusivity of a dilute suspension red blood cells to find a monotonically decreasing value for $f_{s,2}$. Note however that the self and gradient diffusivities although related describes manifestly different aspects of emulsion behaviors and could have different trends. Although, the current work focuses on collective diffusivity, we compare here our simulation method with that of Loewenberg and Hinch [4]. Computing self-diffusivity by simulating pair-interactions between droplets using their method would require for a single set of parameters ~60-70 simulations with different relative initial positions of the droplets. It would be computationally expensive, onerous and is clearly outside the scope of the present work. However, here we compute pair-interactions for droplets in shear to show in Figure 5(a) that the final net relative displacement (initially at (0,0,0) and (-10$a$,0.5$a$,0)) as a function of viscosity ratio matches identically with the boundary element simulations of Loewenberg and Hinch [4]. In Figure 5(b), the relative $y$-displacement between drops as a function of relative $x$-displacement has been shown for two different capillary numbers to also offer excellent match with experiments [51] as well as more recent boundary element simulation from the same group Cristini, Błwazdziewicz [52]. Successful comparison with previous methods offers validation for our methods, specifically for our numerical simulation technique. It remains difficult to conclusively argue for a simple physical reason for the trend in effective properties of overall emulsions such as gradient- or self-diffusivities. However, the monotonic decrease of final relative displacement in

Figure 6(a) offers a plausible reason for monotonically decreasing self-diffusivity in a dilute emulsion, i.e., neglecting more-than-two-particle interaction, as seen in Loewenberg and Hinch [4].

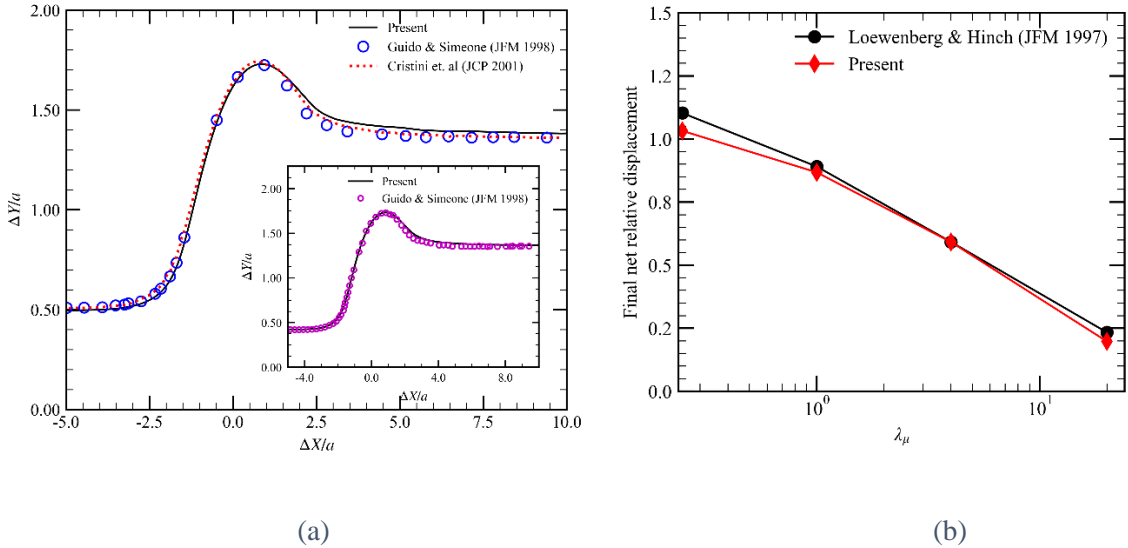

(a)　　　　　　　　　　　　　　　　(b)

*Figure 5 (a) Relative displacement of two drops interacting in shear flow comparing our results (solid line) with those obtained from boundary element method (dotted line) by Cristini, Bławzdziewicz [52] and experimental measurements(open circles) by S. and M. [53]. Main figure is with Ca=0.135, $\lambda_\mu = 1.37$. Inset is with Ca=0.13 and $\lambda_\mu = 1.4$.(b) Final net relative displacement of two drops, initially separated 0.5a in the gradient direction, for different viscosity ratios, $\lambda_\mu$, at Ca=0.30. Present results(diamonds) compare reasonably well with boundary element method results of Loewenberg and Hinch [4].*

### 3.1.2　$D^c_{yy}/\dot{\gamma}a^2$ using dynamic structure factor

We compute $F(\mathbf{k},t')$ using (6) from the drop positions (discarding initial data $t'=20$). The resulting intermediate scattering functions are averaged over overlapping intervals to obtain a smooth time evolution curve [14]. Figure 6(a) shows that $-\ln F(\mathbf{k}',t)/k^2$ at $Ca$ =0.20 and $\lambda_\mu$ =5.0 for different wavenumbers normalized by their initial value displays a linear growth with the slope being asymptotic to a single value in the limit of $k\to 0$ or the large wavelength limit. This limiting slope, $D_{c,yy}$, is plotted as a function of $\lambda_\mu$ for several Ca values in Figure 6(b). The trend is remarkably similar to that of $f_2$ vs $\lambda_\mu$ in Figure 4(a). A direct comparison is not available as the emulsion is not homogeneous and the average concentration of the progressively widening drop layer decreases with time. However note that the ratio $D^c_{yy}/f_2$ ~0.1, is close to an average volume fraction $\varphi$ over the whole diffusive process starting with the maximum $\varphi_{initial}$

~0.25 for the initial packed layer. We plot in the inset of Figure 6(a) the slope of $-\ln F / k^2$ as a function of wavenumber k, for different $\lambda_\mu$. However a straightforward interpretation of this quantity as the wave number dependent diffusivity is not possible as is done in the literature [14] because of the inhomogeneous nature of the problem. Here the process is dominated by an initial inhomogeneous distribution containing finite concentration gradient relaxing through a nonlinear diffusive process (Eq.(1)) unlike the typical case in the literature where it is really a spontaneously arising infinitesimal stochastic concentration fluctuation relaxing. As a result, the slope increases with $k$ unlike say in Leshansky and Brady [14] where it decreases reaching self-diffusivity in the limit of $k \to \infty$. However, we note that the dynamic structure factor, which were typically computed and applied only to homogeneous systems, offers a novel perspective about gradient diffusivity of non-homogeneous systems. Finally, the remarkable similarity of the curves in Figures 5(a) and 7(b), despite completely independent way of computing them makes us confident about the trends of variations of $f_2$ as well as $D^c_{yy}$.

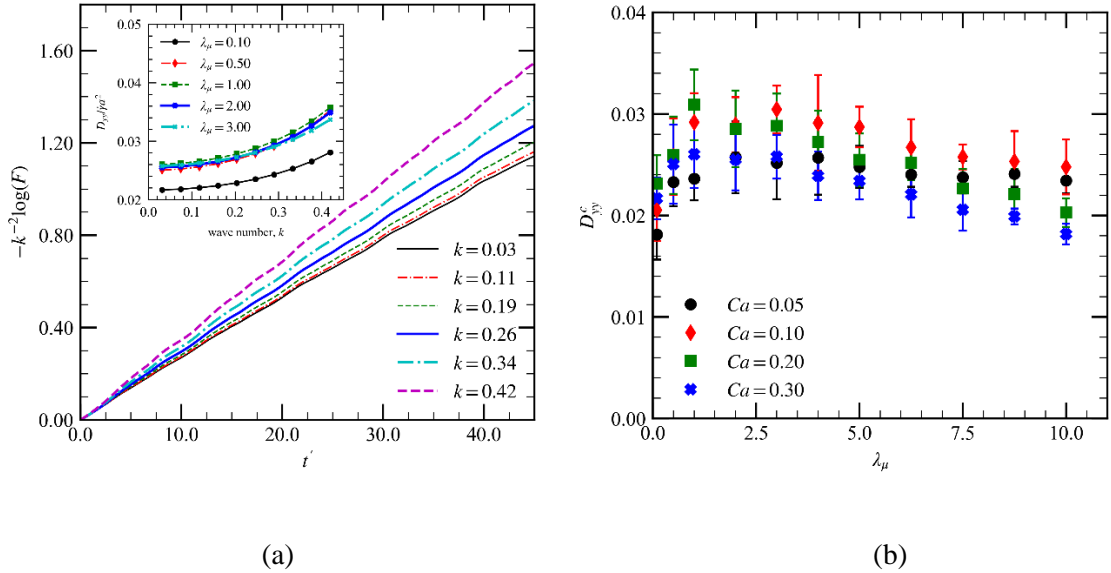

(a)          (b)

*Figure 6 (a) $D^c_{yy} / \dot{\gamma} a^2$ vs wavenumber,k for a few different viscosity ratios with Ca=0.20 and $\lambda_\mu$ =5.0 (b) $D^c_{yy} / \dot{\gamma} a^2$ vs. viscosity ratio for different Ca.*

## 3.2 Effects of Capillary number variation

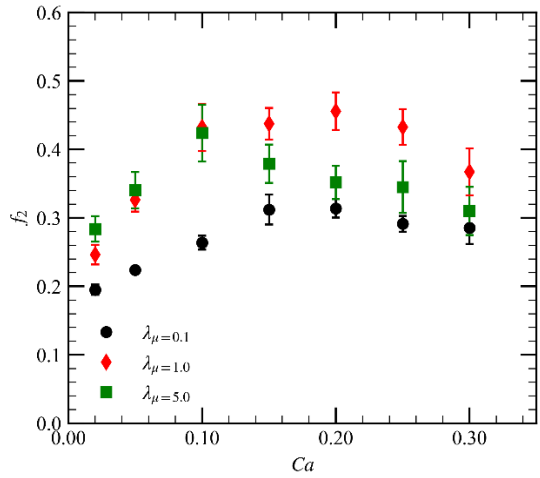
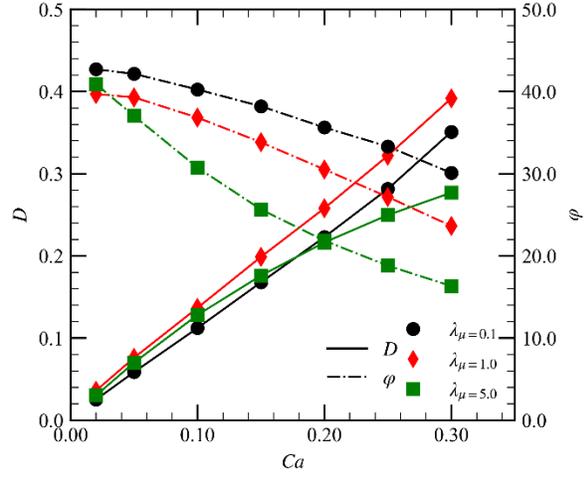

(a)

(b)

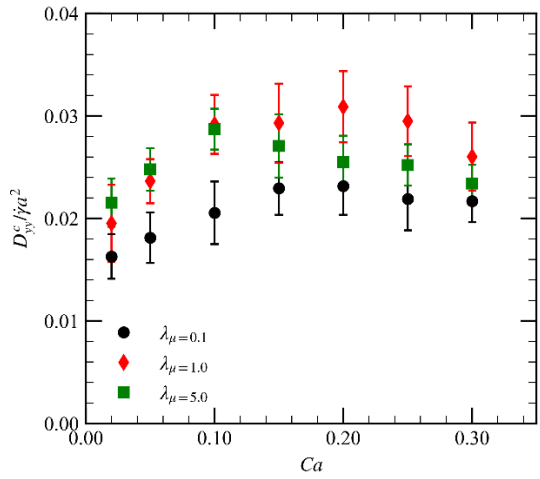
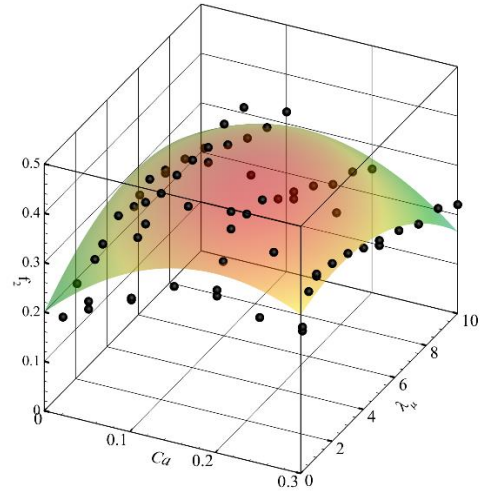

(c)

(d)

*Figure 7 (a) Dimensionless gradient diffusivity $f_2$ and (b) average drop deformation parameter, orientation angle vs. Ca for different viscosity ratios. (c) $D^c_{yy}/\dot{\gamma}a^2$ vs Ca for different viscosity ratio. (d) Surface plot showing $f_2$ as a function of capillary number, Ca, and viscosity ratio, $\lambda_\mu$. The correlation describing the surface is given in Eq. 9.*

Figure 7(a) plots $f_2$ vs. *Ca* for three different viscosity ratios, all showing the nonmonotonic variation. As noted in [8], the nonmonotonicity arises from the competition between rising deformation and decreasing inclination with increasing *Ca* (shown in Figure 7(b)), the former increasing diffusivity initially, and the latter eventually dominating to decrease it. Note that as we noted in [8], slightly different explanation was offered by Loewenberg and Hinch [4] for the nonmonotonic variation of the self-diffusivity coefficient $f_{s,2}$ with *Ca*. In Figure 7(c), we also plot $D_{yy}^c$ as a function of *Ca* for the same three $\lambda_\mu$ values. It shows again very similar trends as in Figure 7(c). Note that the values found here are in agreement with the experimental measurement of Hudson [17] who performed the experiments for viscosity ratios of 0.17-0.19 (depending on the drop liquids) obtaining $f_2$ = 0.16-0.25 for *Ca* of 0.02-0.40. Grandchamp et. al. [6] measured a value of ~0.77 (rescaled by particle volume) for red blood cells.

Finally, by combining all the results from the previous sections, we have generated an empirical correlation for the dimensionless diffusivity, $f_2$, as a function of capillary number and viscosity ratio:

$$f_2 = -4.06 Ca^2 + 3.12 \times 10^{-3} \lambda_\mu^2 - 6.52 \times 10^{-2} Ca \cdot \lambda_\mu \\ + 1.62 Ca + 3.52 \times 10^{-2} \lambda_\mu + 0.2. \tag{11}$$

The empirical correlation is plotted in Figure 7(d). Similarly, an empirical correlation for $D_{yy}^c / \dot\gamma a^2$ is also obtained (not plotted as the variation is similar to Figure 7(d)):

$$D_{yy}^c / \dot\gamma a^2 = (-2.36 Ca^2 - 1.67 \times 10^{-3} \lambda_\mu^2 - 3.88 \times 10^{-2} Ca \cdot \lambda_\mu \\ + 0.94 Ca + 1.86 \times 10^{-2} \lambda_\mu + 0.17)/10. \tag{12}$$

## 4   Conclusion

We have computed the shear induced gradient diffusivity in a sheared viscous emulsion of droplets using a front-tracking based direct numerical simulation. We simulated the deformation and motion of droplets initially packed in a layer subjected to a simple shear. We focused on the effects of the varying viscosity ratio between the drop and matrix fluids. From the time evolution of the droplet phase concentration, we compute the coefficient $f_2$ of the gradient or collective diffusivity using a self-similar solution of the one dimensional nonlinear diffusion equation assumed to be satisfied by the system. There haven't been any prior numerical computation of the quantity in the literature for emulsion of viscous drops, capsules or vesicles. There has been one experimental measurement of this quantity for viscous emulsion in the literature, which is in agreement with the values computed here. It is also of the same order as what has been measured in suspensions of cells and vesicles.

The computed coefficient varied non-monotonically with the viscosity ratio—in the range of viscosity ratio [0.1, ~2] gradient diffusivity increased with viscosity ratio and decreases for values beyond this range. The nonmonotonic behavior arises due to nonmonotonic variation in drop deformation with viscosity ratio. The nonmonotonic trend is slightly different from the one computed for self- diffusivity computed in the literature using pair-interactions for both viscous droplets [4]—the self-diffusivity coefficient remained constant for smaller viscosity ratios decreasing monotonically at higher viscosity ratios—and capsules [23]—monotonically decreasing with the viscosity ratio. We validated our computational technique against boundary element technique used by those authors. This led us to believe that the non-monotonic behavior results from multi-particle interactions, or self and gradient diffusivities have different variations with viscosity ratio at low viscosity ratio, an issue to be explored in future work.

We also compute the gradient diffusivity $D_{yy}^c$ computing the dynamic structure factor, a computation that has so far been restricted to homogeneous suspension. The results when appropriately scaled with average volume fraction are in agreement with the computed $f_2$ values, and more importantly they show very similar variation with viscosity ratio, although the two methodologies are very different. Based on the simulation results, we offer two phenomenological correlations for $f_2$ and $D_{yy}^c$ as functions of capillary number and viscosity ratio.

**Acknowledgement**

KS acknowledges partial support from George Washington University. Authors acknowledge time on Colonial One cluster at GWU. This also work used the Comet cluster at the at the San Diego Supercomputer Center, through the Extreme Science and Engineering Discovery Environment (XSEDE) program which is supported by National Science Foundation grant number ACI-1548562.